\documentclass{svjour3}                     
\smartqed  
\usepackage{graphicx,url}
\usepackage{amsmath,amssymb}
\begin{document}

\title{The scalar field source in Kaluza-Klein theory
}


\author{Paul S. Wesson         \and
        James M. Overduin 
}

\institute{P.S. Wesson \at
              Department of Physics and Astronomy, University of Waterloo\\
Waterloo, Ontario, N2L 3G1, Canada \\
              \email{psw.papers@yahoo.ca}           
           \and
           J.M. Overduin \at
              Department of Physics, Astronomy \& Geosciences, Towson University\\
Towson, Maryland 21252, U.S.A.\\
	      \email{joverduin@towson.edu}
}

\date{Received: date / Accepted: date}

\maketitle

\begin{abstract}
To better understand the scalar field typical of higher-dimensional extensions of general relativity, we analyse three classes of solutions.
In all, the field equation for the extra dimension resembles the Klein-Gordon equation, and we evaluate the strength of the source.
Our results show that the scalar field is coupled to matter, and may be regarded as generating it.
\end{abstract}

\maketitle

\section{Introduction}	

The extension of Einstein's general relativity to more than four spacetime dimensions is widely seen as one of the most promising routes to unification of the fundamental interactions \cite{OW97}.
Questions have persisted, though, about the role of the scalar fields (sometimes referred to as dilatons) that are generically associated with each additional dimension.
In compactified Kaluza-Klein theories, non-observation of these fields is traditionally explained by quantum effects.
Here we wish to take a different approach and investigate the properties of the Kaluza-Klein scalar when compactification is not necessarily imposed from the outset.

We take the five-dimensional field equations as expressed in four dimensions \cite{WP92}, and analyse the relation for the scalar field in three classes of exact solutions \cite{P88,WLS00,WO12}.
In all three cases, we find that the field equation resembles the Klein-Gordon equation, with a source strength that depends on the properties of matter.
We are led to the tentative conclusion that the scalar field couples to matter in a way that is analogous to the coupling between the gravitational field and mass, or between the electromagnetic field and charge.

These results, which have not been discussed elsewhere, are of physical interest because they may provide ways to interpolate between classical and quantum approaches to matter.
For example, the second class of solutions considered below describes the birth of a uniform universe from the penetration of an extra-dimensional wave into spacetime, reminiscent of scenarios in which the big bang is associated with a quantum-mechanical tunnelling event \cite{V82}.

However, the mathematical work required to isolate the scalar sources in all three classes of solutions is lengthy, and we will generally present only the final results below.
Our notation is standard: upper-case Latin letters read 0,123,4 while lower-case Greek letters read 0,123 and the fundamental constants are usually absorbed.

\section{The 5D Field Equations} \label{pw-sec2}

There are many forms of the field equations and numerous exact solutions of them are known \cite{OW97}.
In general, there are 15 such equations, which serve to determine the 10 gravitational potentials, the 4 for electromagnetism, and the 1 for the scalar field $\Phi$.
The field equation for the last is the 44 component.
To study the relationship between $\Phi$ and the properties of matter, it is necessary to formulate the 5D field equations in a form which includes the 4D Einstein tensor, which can be matched to an energy-momentum tensor.
The way to do this has been known for some time \cite{WP92}, and is followed here.

The coordinates are taken to be $x^A=x^{\gamma},\ell$ where $x^{\gamma}=0,123$ refer to time and space, while $x^4=\ell$ refers to the extra dimension.
The field equations for 5D relativity are commonly taken in terms of the Ricci tensor as
\begin{equation}
R_{AB} = 0 \;\;\; (A,B = 0,123,4) \, .
\label{pw-eq1}
\end{equation}
To solve these equations, the 5D metric can be simplified by using 4 of the 5 available degrees of coordinate freedom to set the potentials of electromagnetic type to zero, via $g_{4\alpha}=0$.
The remaining degree of coordinate freedom is sometimes used to set the scalar field to a constant, via $|g_{44}|=1$.
But to bring out the effects of this field, we instead specify it by $g_{44}=\varepsilon\Phi^2$, where $\Phi=\Phi(x^{\alpha},\ell)$ and $\varepsilon=\pm 1$ allows for both a spacelike and timelike extra dimension.
(The extra dimension for $\varepsilon=+1$ need not have the physical nature of a time so there is no problem with closed historical paths.) The 5D line element then takes the form
\begin{equation}
dS^2 = g_{\alpha\beta}(x^{\gamma},\ell)dx^{\alpha}dx^{\beta} + \varepsilon\Phi^2(x^{\gamma},\ell)d\ell^2 \, .
\label{pw-eq2}
\end{equation}
This includes the 4D line element $ds^2=g_{\alpha\beta}dx^{\alpha}dx^{\beta}$, and allows the use of the 4D proper time $s$ as a parameter in the study of dynamics (see below).
With the metric~(\ref{pw-eq2}), the field equations~(\ref{pw-eq1}) can be conveniently grouped into sets of 10 (tensor), 4 (vector) and 1 (scalar), thus:
\begin{eqnarray}
G_{\alpha\beta} & = & 8\pi T_{\alpha\beta} \nonumber \\
8\pi T_{\alpha\beta} & \equiv & \frac{\Phi_{,\alpha;\beta}}{\Phi}-\frac{\varepsilon}{2\Phi^2}\left\{\frac{\Phi_{,4}\,g_{\alpha\beta,4}}{\Phi}-g_{\alpha\beta,44}+g^{\lambda\mu}g_{\alpha\lambda,4}\,g_{\beta\mu,4} \right. \nonumber \\
& & -\left. \frac{g^{\mu\nu}g_{\mu\nu,4}\,g_{\alpha\beta,4}}{2} + \frac{g_{\alpha\beta}}{4}\left[g^{\mu\nu}_{,4}\,g_{\mu\nu,4}+\left(g^{\mu\nu}g_{\mu\nu,4}\right)^2\right]\right\} 
\label{pw-eq3}\\
P^{\beta}_{\alpha;\beta} & = & 0 \nonumber \\
P^{\beta}_{\alpha} & \equiv & \frac{1}{2\Phi} \left(g^{\beta\sigma}g_{\sigma\alpha,4} - \delta^{\beta}_{\alpha}\,g^{\mu\nu}g_{\mu\nu,4}\right) 
\label{pw-eq4} \\
\Box\Phi & = & -\frac{\varepsilon}{2\Phi} \left[ \frac{g^{\lambda\beta}_{,4}g_{\lambda\beta,4}}{2}+g^{\lambda\beta}g_{\lambda\beta,44}-\frac{\Phi_{,4}\,g^{\lambda\beta}g_{\lambda\beta,4}}{\Phi} \right] \nonumber \\
\Box\Phi & \equiv & g^{\alpha\beta}\Phi_{,\alpha;\beta}
\label{pw-eq5} \, .
\end{eqnarray}
Here a comma denotes the partial derivative, and a semicolon denotes the standard (4D) covariant derivative.
These equations are algebraically general, and can be applied to any physical problem where gravitational and scalar fields are dominant.

In what follows, we present three classes of solutions to Eqs.~(\ref{pw-eq3})-(\ref{pw-eq5}), express the energy-momentum tensor in the form of a perfect fluid, $T_{\alpha\beta}=(\rho+p)u_{\alpha}u_{\beta}+pg_{\alpha\beta}$, and rewrite the scalar-field relation~(\ref{pw-eq5}) in terms of the properties of matter.
In this way, we will obtain relations $\Phi=\Phi(\rho,p)$ which show the connection between the scalar field and the density and pressure of matter.
As far as we are aware, this has not been done before (even though the first class of solutions has been known for a long while).
Before proceeding, we note that it is not possible to transpose Eqs.~(\ref{pw-eq3})-(\ref{pw-eq5}) in such a manner as to obtain a generic relation of the form $\Phi=\Phi(\rho,p)$.
It is necessary to analyse each solution independently.
The required working is long, even though the solutions themselves may be verified quickly by computer \cite{OE13}.

\section{5D Friedmann-Robertson-Walker Cosmologies} \label{pw-sec3}

These models were found originally by Ponce de Leon \cite{P88}, and have been much studied.
The 5D metric reduces to the 4D Robertson-Walker one of standard cosmology on the hypersurfaces $x^4=\ell=$constant.
The 3D sections are flat, but after their discovery it was also found that they are 5D flat.
Indeed, it is now acknowledged that the 5D analogs of all FRW models are flat \cite{L00}.
They are, however, curved in 4D and therefore contain matter with density $\rho$ and pressure $p$, which can be found by evaluating (\ref{pw-eq3}) for the effective energy-momentum tensor.

The line element is given by
\begin{equation}
dS^2 = \ell^2dt^2-t^{2/\alpha}\ell^{2/(1-\alpha)}\left(dr^2+r^2d\Omega^2\right)-\frac{\alpha^2t^2}{(1-\alpha)^2}d\ell^2 \, ,
\label{pw-eq6}
\end{equation}
where $d\Omega^2\equiv(d\theta^2+\sin^2\theta d\phi^2)$.
The dimensionless parameter $\alpha$ is related to the properties of matter.
These can be obtained using the technique outlined above, giving from (\ref{pw-eq3}):
\begin{equation}
8\pi\rho = \frac{3}{\alpha^2\tau^2} \, , \hspace{1cm} 8\pi p = \frac{2\alpha-3}{\alpha^2\tau^2} \, .
\label{pw-eq7}
\end{equation}
Here $\tau=\ell\,t$ is the proper time.
The equation of state is $p=(2\alpha/3-1)\rho$.
For $\alpha=3/2$, the scale factor of Eq.~(\ref{pw-eq6}) varies as $t^{2/3}$, the density and pressure of Eq.~(\ref{pw-eq7}) are $\rho=1/6\pi\tau^2$ with $p=0$, and we have the standard dust-dominated model.
For $\alpha=2$, the scale factor varies as $t^{1/2}$, $\rho=3/32\pi\tau^2=3p$ and we have the standard radiation model for the early universe.
(The late vacuum-dominated era is formally approached as $\alpha\rightarrow0$.)

Turning our attention to Eqs.~(\ref{pw-eq4}) and (\ref{pw-eq5}), we can now evaluate the components of the 4-tensor $P^{\beta}_{\alpha}$ and express the scalar field $\Phi$ as a function of the properties of matter.
For $P^{\beta}_{\alpha}$ we find:
\begin{equation}
P_0^0 = -3/\alpha T \, , \hspace{1cm} P_1^1 = P_2^2 = P_3^3 = (\alpha-3)/\alpha T \, .
\label{pw-eq8}
\end{equation}
In terms of these, the physical density and pressure are
\begin{equation}
8\pi\rho = (P_0^0)^2/3 \, , \hspace{1cm} 8\pi p = P_0^0 ( P_0^0 - 2P_1^1)/3 \, .
\label{pw-eq9}
\end{equation}
For $\Phi$ we find:
\begin{equation}
\Box\Phi = (3/2)8\pi(\rho+p)\Phi \, .
\label{pw-eq10}
\end{equation}
This reveals that the evolution of the scalar field depends on what is sometimes called the inertial mass density $\rho+p$, a name which follows from the fact that in FRW models the rate of change of the density is proportional to this combination, which thereby determines the stability of matter.
It is important to realize that the field equations~(\ref{pw-eq4}) and (\ref{pw-eq5}) for $P^{\beta}_{\alpha}$ and $\Phi$ have no counterparts in 4D general relativity, so their physical interpretation is to a certain extent open.
For the present class of solutions~(\ref{pw-eq6}), the $P^{\beta}_{\alpha}$ of Eq.~(\ref{pw-eq8}) appear to be matter currents, while the $\Phi$ of Eq.~(\ref{pw-eq10}) is some kind of matter field.

\section{5D Wave-like Cosmologies}

These models were found by Wesson, Liu and Seahra \cite{WLS00}, but their detailed properties have not been studied.
They were found originally during a study of metrics with wave-like properties, where the essential quantities depend on the combined variable $u\equiv(t-\ell)$.
However, as will be seen below, they change under a transformation of coordinates to FRW-like cosmologies, with flat 3D sections like the solutions treated above.

The line element is given by
\begin{eqnarray}
& & dS^2 = b^2 dt^2 - a^2 \left( dr^2 + r^2 d\Omega^2 \right) - b^2 d\ell^2 \nonumber \\
& & a = (hu)^{1/(2+3\alpha)} \, , \hspace{1cm} b = (hu)^{-(1+3\alpha)/2(2+3\alpha)} \, .
\label{pw-eq11}
\end{eqnarray}
There are two arbritrary constants.
The first ($h$) is kept separate from the dynamical variable $u\equiv(t-\ell)$ because, as will become apparent below, it has the physical dimensions of an inverse length or time, and is related to Hubble's parameter.
The second constant ($\alpha$) turns out to be related to the properties of matter, as in the previous class of solutions, though it is not identical to the one used before.
In fact, by the Einstein-like set of field equations~(\ref{pw-eq3}) matched to a perfect-fluid energy-momentum tensor, we find
\begin{eqnarray}
& & p = \alpha\rho \nonumber \\
& & 8\pi\rho = \frac{3h^2}{(2+3\alpha)^2} a^{-3(1+\alpha)} \, .
\label{pw-eq12}
\end{eqnarray}
We see that the equation of state is isothermal, where $\alpha=0$ corresponds to the pressureless (dust) universe and $\alpha=1/3$ corresponds to the early (radiation) universe.
(The late vacuum-dominated era is formally approached as $\alpha\rightarrow -1$.) To further elucidate the physical properties of the solution, it is instructive to change from the coordinate time $t$ to the proper time $T$.
This is defined by $dT=b\,dt$, so 
\begin{equation}
T = \frac{2}{3} \left( \frac{2+3\alpha}{1+\alpha}\right)\frac{1}{h}(hu)^{3(1+\alpha)/2(2+3\alpha)} \, .
\label{pw-eq13}
\end{equation}
The 4D scale factor which determines the dynamics of the model by Eqs.~(\ref{pw-eq11}) and (\ref{pw-eq13}) is then
\begin{equation}
a(T) = \left[ \frac{3}{2} \left( \frac{1+\alpha}{2+3\alpha}\right)hT\right]^{\frac{2}{3(1+\alpha)}} \, .
\end{equation}
Fo $\alpha=0$, $a(T)\sim T^{2/3}$ as in the (Einstein-de~Sitter) dust model.
For $\alpha=1/3$, $a(T)\sim T^{1/2}$ as in the standard radiation model.
The value of Hubble's parameter is given by
\begin{eqnarray}
H & \equiv & \frac{1}{a}\frac{\partial a}{\partial T} = \frac{h}{(2+3\alpha)} (hu)^{-3(1+\alpha)/2(2+3\alpha)} \nonumber \\
& = & \frac{2}{3(1+\alpha)T} \, .
\end{eqnarray}
This shows that for $\alpha=0$ and $1/3$, $H$ has its standard values in terms of the proper time.
We can also convert the density~(\ref{pw-eq12}) from $t$ to $T$ using Eq.~(\ref{pw-eq13}) and find
\begin{equation}
8\pi\rho = \frac{4}{3} \frac{1}{(1+\alpha)^2} \frac{1}{T^2} \, .
\label{pw-eq16}
\end{equation}
For $\alpha=0$ we have $\rho=1/6\pi T^2$, and for $\alpha=1/3$ we have $\rho=3/32\pi T^2$, the standard FRW values.
Thus, the 5D solution~(\ref{pw-eq11}) contains 4D dynamics and matter which are the same as in the standard 4D cosmologies.
However, the interpretation of the big bang is quite different, depending on whether the viewpoint is 4D or 5D.
The big bang occurs in proper time at $T=0$ by Eq.~(\ref{pw-eq16}), but it occurs in coordinate time at $a=0$ or $u=(t-\ell)=0$ by Eqs.~(\ref{pw-eq12}) and (\ref{pw-eq11}).
Therefore, we can interpret the big bang either as a singularity in 4D or as a hypersurface $t=\ell$ that represents a plane wave propagating in 5D.

We can gain further insight by using the solutions~(\ref{pw-eq11}) with the purely extra-dimensional field equations~(\ref{pw-eq4}) and (\ref{pw-eq5}).
Of particular interest is the scalar $P$, which is the sum of all the components in Eq.~(\ref{pw-eq4}), and its square.
These can be expressed either in terms of the original combined variable $u\equiv(t-\ell)$ of Eq.~(\ref{pw-eq11}), or in terms of the density $\rho$ of Eq.~(\ref{pw-eq12}).
We find:
\begin{eqnarray}
P & = & -\frac{2(3\alpha-5)}{2(2+3\alpha)} \frac{1}{bu}
\label{pw-eq17} \\
P^2 & = & \frac{3}{4} (3\alpha-5)^2 8\pi\rho \, .
\label{pw-eq18}
\end{eqnarray}
Insofar as $P^2$ is proportional to the matter density, there may appear to be some justification for the practice of regarding this quantity as a contribution to the total density by the singular hypersurface of membrane theory \cite{OW97}.
However, when the 5D field equations are expressed as in Section~\ref{pw-sec2}, it is seen that the tensor $P^{\beta}_{\alpha}$ is exactly conserved {\it by itself}, as in Eq.~(\ref{pw-eq4}).
This implies that the components of $P^{\beta}_{\alpha}$ should rather be regarded as matter currents, in the manner of Section~\ref{pw-sec3}.
There, we also found that the field equation $R_{44}=0$, which has the algebraic form of a wave equation for the scalar field $\Phi$, could be regarded physically as a kind of generation relation for the matter.
That interpretation is supported here, because we find that Eq.~(\ref{pw-eq5}) can be written in the following alternative forms:
\begin{eqnarray}
\Box\Phi & = & \frac{(1+3\alpha)(3\alpha-1)}{2(2+3\alpha)^2} \frac{1}{\Phi u^2} 
\label{pw-eq19} \\
& = & \frac{(1+3\alpha)(3\alpha-1)}{6} \, 8\pi\rho\,\Phi \, .
\label{pw-eq20}
\end{eqnarray}
This last relation bears some resemblance to Eq.~(\ref{pw-eq10}) of Section~\ref{pw-sec3}, and we will discuss both of them in Section~\ref{pw-sec6}.

Before that, we wish to look at another approach to the field equations, namely by using a gauge.
This basically involves the use of a set of coordinates which is suggested by physics but leads to a simplification of the field equations.
There are an infinite number of possible gauges, but only two are in widespread use.
Both employ a function of the extra coordinate $(x^4=\ell\,)$, which is multiplied onto the 4D part of the metric, while the fifth part is flat $(\Phi=1)$.
In membrane theory, mentioned above, the interactions of particles are concentrated around a singluar hypersurface, which is identified with spacetime, and to achieve this the gauge function is taken to be an exponential, resulting in what is commonly called the warp metric.
In Space-Time-Matter theory, which lies behind the form  of the field equation~(\ref{pw-eq3})-(\ref{pw-eq5}), the terms involving the fifth dimension are responsible for matter and vacuum, and these are best handled if the gauge function is taken to be a quadratic, resulting in what is commonly called the canonical metric.
Unfortunately, both of these gauges fail to bring out the effects of the scalar potential, so we turn our attention to a new and broader one.

\section{5D Scalar-Field Models}

The object of the exercise is to restrict the coefficients in the general metric~(\ref{pw-eq2}) in such a way that the field equations~(\ref{pw-eq3})-(\ref{pw-eq5}) become tractable but yield significant results, particularly concerning the scalar field $\Phi$.
Consider, therefore, the following:
\begin{equation}
g_{\alpha\beta}(x^{\gamma},\ell\,) = \exp(\ell\,\Phi/L)\,\bar{g}_{\alpha\beta}(x^{\gamma}) \, , \hspace{1cm} \Phi=\Phi(x^{\gamma}) \, .
\label{pw-eq21}
\end{equation}
Here $L$ is a constant length whose meaning will shortly become clear.
This gauge, when substituted into the field equations~(\ref{pw-eq3})-(\ref{pw-eq5}), yields some interesting results.

The effective energy-momentum tensor~(\ref{pw-eq3}) is given by
\begin{equation}
8\pi T_{\alpha\beta} = \frac{\Phi_{,\alpha;\beta}}{\Phi}-\frac{\varepsilon}{2L^2} g_{\alpha\beta} \, ,
\label{pw-eq22}
\end{equation}
where we drop the bar on $\bar{g}_{\alpha\beta}$ for clarity.
The second term here is typical of a 4D Einstein space with cosmological constant $\Lambda=-\varepsilon/2L^2$.
This is positive for a spacelike extra dimension $(\varepsilon=-1)$ and negative for a timelike one $(\varepsilon=+1)$.
The trace of Eq.~(\ref{pw-eq22}) is
\begin{equation}
8\pi T = \frac{\Box\Phi}{\Phi} - \frac{2\varepsilon}{L^2} \, ,
\label{pw-eq23}
\end{equation}
which will be used below.

The tensor defined by Eq.~(\ref{pw-eq4}) and its associated scalar are given by
\begin{eqnarray}
& & P_{\alpha}^{\beta} = -\frac{3}{2L} \delta_{\alpha}^{\beta} \\
& & P = -6/L \, .
\end{eqnarray}
Since the 4-tensor $P_{\alpha}^{\beta}$ is conserved, $P$ is obviously a constant for the spacetime.
It is possible to give a physical interpretation of $P_{\alpha}^{\beta}$ , in a manner analogous to how the $T_{\alpha\beta}$ of Eq.~(\ref{pw-eq3}) is written in terms of the properties of matter.
In both cases, the form depends on whether the source is a fluid or a discrete particle.
For a fluid, the physical form of $P$ and $P^2$ were given above for the metric~(\ref{pw-eq11}) as Eqs.~(\ref{pw-eq17}) and (\ref{pw-eq18}).
In the present case, the gauge~(\ref{pw-eq21}) is particularly interesting when interpreted as a particle moving through a vacuum \cite{WO12}.
Then a reasonable physical form is $P_{\alpha\beta} = (mc/h)u_{\alpha}u_{\beta}$ in terms of the mass measured in atomic units and the 4-velocities ($h$ is Planck's constant and $c$ is the speed of light).
This prescription means that the field equations $P_{\alpha;\beta}^{\beta}=0$ give back the usual geodesic equations of motion, $u^{\alpha}_{;\beta}u^{\beta}=0$, and the conserved quantity is then just the mass of the particle.

The scalar equation~(\ref{pw-eq5}) can be written
\begin{equation}
\Box\Phi + \frac{\varepsilon}{L^2}\Phi = 0 \, .
\label{pw-eq26}
\end{equation}
This relation has many well-known applications in physics, which we will summarize in Section~\ref{pw-sec6}.
When $\Phi$ depends on all four spacetime coordinates, it is formally identical with the Klein-Gordon equation (if $\varepsilon=+1$).
The latter is the wave equation for a relativistic particle of rest mass $m$, where $L$ is then the Compton wavelength of the particle, $h/mc$.
It should be noted that for the 5D canonical gauge, it is possible to recover the wave function if the extra dimension is timelike, in which case the Klein-Gordon equation is obtained from the extra component of the geodesic equation \cite{W11}.
For the new gauge~(\ref{pw-eq21}), the real or complex nature of the scalar field $\Phi$ is not specified, and its physical interpretation depends partly on $\varepsilon=\pm1$.
(In this regard, $\ell\rightarrow i\ell$ also formally changes the sign of the last term in the 5D metric.) It is due to the fact that $\varepsilon$ can in principle have either sign that it is kept explicit in the present account.
This choice affects not only the interpretation of the scalar field equation~(\ref{pw-eq26}) but also the sign of the 4D scalar curvature.
This is given in general and for the gauge~(\ref{pw-eq21}) by
\begin{equation}
R = \frac{\varepsilon}{4\Phi^2}\left[g^{\mu\nu}_{,4}\,g_{\mu\nu,4} + \left(g^{\mu\nu}\,g_{\mu\nu,4}\right)^2\right] = \frac{3\varepsilon}{L^2} \, .
\end{equation}
This may be compared to Eq.~(\ref{pw-eq23}), which with Eq.~(\ref{pw-eq26}) gives the trace of the Einstein tensor as $G=-R$, as expected.
The dependence of the 4D curvature on $1/\Phi^2$, as shown here, is typical of how 4D quantities scale in regard to the strength of the scalar field.
For this reason, $\Phi$ is sometimes called the inflaton \cite{OW97}.
However it is clear from what has been shown above that its effects are actually quite diverse.
To reveal further effects of $\Phi$, especially in quantum systems governed by the Klein-Gordon equation, more work should be done on the gauge outlined in this section.
For example, in Eq.~(\ref{pw-eq21}) the exponent could be changed from $\Phi(x^{\gamma})\,\ell$ to $\int\Phi(x^{\gamma},\ell)\,d\ell$, which would considerably enlarge the scope of the resulting physics.

\section{Discussion and Conclusion} \label{pw-sec6}

Kaluza-Klein theory in its general form is a unified account of gravitational, electromagnetic and scalar fields, whose potentials are represented by the $g_{\alpha\beta}$, $g_{4\alpha}$ and $g_{44}$ components of the 5D metric tensor.
The 5D field equations~(\ref{pw-eq1}) are expected to contain the Einstein field equations, the Maxwell equations and the Klein-Gordon equation.
However, because the theory is covariant, these fields may be mixed, and appear in guises that depend on the choice of coordinates (or gauge).
In the present account, the $g_{4\alpha}$ potentials were eliminated by the choice of the starting metric~(\ref{pw-eq2}) so as to concentrate on the effects of the gravitational and scalar fields.
These effects were studied for three classes of 5D solutions: FRW cosmologies with the metric~(\ref{pw-eq6}), wave-like cosmologies with the metric~(\ref{pw-eq11}), and models dominated by the scalar field with metric~(\ref{pw-eq21}).
Because they are poorly understood, particular attention was paid to the conserved 4-tensor $P_{\alpha\beta}$ and the scalar field $\Phi$, both of which lie outside 4D general relativity.
Somewhat surprisingly, the equation governing $\Phi$ was found to have the same form in all three cases, given by Eqs.~(\ref{pw-eq10}), (\ref{pw-eq20}) and (\ref{pw-eq26}).
The last of these illustrated the generic form: $\Box\Phi + \varepsilon\Phi/L^2=0$, where $\varepsilon=\pm1$ and $L$ is a physical length which incorporates the properties of the source.

The noted relation is, of course, common in the literature of physics.
In classical physics, it is known as the Helmholtz equation if $\varepsilon=+1$ and the (space part of the) diffusion equation if $\varepsilon=-1$.
In quantum physics, it is known as the Klein-Gordon equation, where $\varepsilon=+1$ and $L$ is the Compton wavelength $(h/mc)$ of the particle; and in the non-relativistic limit this becomes the Schrodinger equation with the kinetic energy as source.
Obviously, for $L\rightarrow\infty$ the source goes away, leaving just Laplace's equation.
The latter is frequently solved by separation of variables, and the same method can be applied to the general equation (which we continue to refer to as Klein-Gordon).
In 5D relativity, covariance means that the coordinates can be chosen at will, and it is often the case that the field equations can be solved much more easily in one set of coordinates than another.
Eisenhart \cite{E34} showed that an equation like $\Box\Phi+\varepsilon\Phi/L^2=0$ can be separated in(at least) eleven coordinate systems.
However, it should be recalled that the basic field equation for the scalar field is $R_{44}=0$ or Eq.~(\ref{pw-eq5}) in expanded form, and it is not known if this can {\it always\/} be written in the form of the Klein-Gordon equation.

Even when the scalar field {\it does\/} take on the form of the Klein-Gordon equation, its solutions differ drastically depending on the choice for $\varepsilon=\pm1$.
This determines whether the extra dimension of 5D relativity is spacelike $(\varepsilon=-1)$ or timelike $(\varepsilon=+1)$.
For the canonical metric mentioned previously, null geodesics in 5D ($dS^2=0$) lead to behaviours of $x^4=\ell$ as a function of 4D proper time $s$ which are either monotonic ($\varepsilon=-1$) or oscillatory ($\varepsilon=+1$).
In fact, it is a general property of solutions of the 5D field equations that those with $\varepsilon=-1$ describe fluid sources which evolve smoothly while those with $\varepsilon=+1$ describe particle-like sources which evolve in a wave-like manner \cite{OW97,W11}.
Also, if there is a vacuum energy density measured by the cosmological constant, this is generally $\Lambda>0$ for $\varepsilon=-1$ and $\Lambda<0$ for $\varepsilon=+1$.
While the form of the field equations precludes a clear-cut division, the evidence suggests that a spacelike extra dimension corresponds to 4D classical behaviour, while a timelike extra dimension corresponds (at least in several cases) to 4D behaviour reminiscent of wave mechanics.

As regards the scalar field, the relation $\Box\Phi+\varepsilon\Phi/L^2=0$ can either be read as the production of the field by the matter {\it or\/} as the production of matter by the field.
The latter interpretation is in accordance with Mach's principle \cite{OF01}.
It might also be seen as a classical analog of the Higgs mechanism in quantum field theory.
In any case, the results noted above show that the scalar field of 5D relativity is intimately connected to the presence of matter.

\section*{Acknowledgments}

We thank various contributors to the Space-Time-Matter consortium (\url{5Dstm.org}) for discussions, and Ryan Everett for help with {\it Mathematica}.

\end{document}